\documentclass[sn-apa]{sn-jnl}% APA Reference Style 
\usepackage{graphicx}%
\usepackage{multirow}%
\usepackage{amsmath,amssymb,amsfonts}%
\usepackage{amsthm}%
\usepackage{mathrsfs}%
\usepackage[title]{appendix}%
\usepackage{xcolor}%
\usepackage{textcomp}%
\usepackage{manyfoot}%
\usepackage{booktabs}%
\usepackage{algorithm}%
\usepackage{algorithmicx}%
\usepackage{algpseudocode}%
\usepackage{listings}%
\usepackage{appendix}
\usepackage[T1]{fontenc}

\theoremstyle{thmstyleone}%
\theoremstyle{thmstyletwo}%

\theoremstyle{thmstylethree}%

\newif\ifarxiv
\arxivtrue
\begin{document}

\title{Prediction-based evaluation of back-four defense with spatial control in soccer}

\ifarxiv
\author[1]{\fnm{Soujanya} \sur{Dash}}

\author[1]{\fnm{Kenjiro} \sur{Ide}}

\author[1]{\fnm{Rikuhei} \sur{Umemoto}}

\author[1]{\fnm{Kai} \sur{Amino}}

\author*[1]{\fnm{Keisuke} \sur{Fujii}}\email{fujii@i.nagoya-u.ac.jp}
% ORCID 0000-0001-5487-4297

\affil[1]{\orgdiv{Graduate School of Informatics}, \orgname{Nagoya University}, \orgaddress{\city{Nagoya}, \country{Japan}}}

\else
Anonymous
\fi 


\abstract{Defensive organization is critical in soccer, particularly during negative transitions when teams are most vulnerable. The back-four defensive line plays a decisive role in preventing goal-scoring opportunities, yet its collective coordination remains difficult to quantify. This study introduces interpretable spatio-temporal indicators namely, space control, stretch index, pressure index, and defensive line height (absolute and relative) to evaluate the effectiveness of the back-four during defensive transitions. Using synchronized tracking and event data from the 2023-24 LaLiga season, 2,413 defensive sequences were analyzed following possession losses by FC Barcelona and Real Madrid CF. Two-way ANOVA revealed significant effects of team, outcome, and their interaction for key indicators, with relative line height showing the strongest association with defensive success. Predictive modeling using XGBoost achieved the highest discriminative performance (ROC AUC: 0.724 for Barcelona, 0.698 for Real Madrid), identifying space score and relative line height as dominant predictors. Comparative analysis revealed distinct team-specific defensive behaviors: Barcelona's success was characterized by higher spatial control and compact line coordination, whereas Real Madrid exhibited more adaptive but less consistent defensive structures. These findings demonstrate the tactical and predictive value of interpretable spatial indicators for quantifying collective defensive performance.
}

% Why did the study need to be done?
% What did you do?
% What did you find?
% How will this study advance the field?
% Paragraph structure
% Brief background (no more than 2 sentence) (why this study need to be done)
% Use signposting (eg, however) to identify research question (why this study need to be done)
% Method/aim (what have you done)
% Result (what you find)
% Conclusion (How it advances the field)




\maketitle

\section{Introduction}
\label{sec:introduction}
Defensive organization is a cornerstone of team success in elite soccer, with the defensive line forming the final outfield barrier against opposing attacks. Among various configurations, the \textit{back four} is a line of four outfield defenders positioned ahead of the goalkeeper which is the most widely adopted system across elite soccer. This configuration is favored by coaches because it provides a clear, stable structure that facilitates rapid reorganization during negative transitions \citep{wright2011role, bradley2011match} where the presence of two central defenders and two fullbacks ensures effective coverage of central and wide areas, reducing the risk of defensive gaps being exploited in moments of instability \citep{casal2016possession}. Compared to back three or back five systems, the back four enables greater tactical flexibility, improved defensive width, and better coordination for pressing and recovery \citep{tierney2013match, tenga2020effect}. Moreover, fullbacks in back four systems are critical for regaining possession and supporting the midfield, further enhancing their team’s ability to quickly re-establish shape after losing the ball \citep{goncalves2023role}. These attributes contribute significantly to the back four’s enduring popularity, especially in the context of elite competition, where the speed and unpredictability of transitions can determine the outcome of matches.

While much of soccer analytics has traditionally emphasized attacking strategies or isolated defensive actions, recent studies underscore the importance of analyzing the defensive line as a coordinated unit. The back four’s synchronization, spacing, and denial of key pitch zones have been identified as central determinants of defensive success, particularly during counterattacks and transitional phases \citep{bojinov2016collective,forcher2022defensive}. Evaluating the collective mechanisms of the defensive line thus provides both theoretical insights into tactical organization and practical applications for coaching interventions. 

Despite this importance, the empirical study of defensive lines has been limited. The complexity of integrating high-resolution tracking and event data, combined with the fluid and reactive nature of defensive behavior, has constrained prior analyses. As the previous work \citep{forcher2022defensive} highlighted in their systematic review, most quantitative research has either focused on broader notions of team compactness or on individual defensive duels, leaving the back four’s coordinated behaviors underexplored. Similarly, another work \citep{casal2021collective} noted that defensive evaluation often relies on outcome-based metrics or manual annotation, which fail to capture dynamic, context-dependent group interactions.

In recent years, efforts have been made to formalize collective defensive performance through interpretable indicators such as \textit{space control}, \textit{pressure}, and the \textit{stretch index}. Space control quantifies the ability of defenders to occupy and deny high-value zones, while pressure captures the intensity and proximity of defensive pressing \citep{taki1996development,Spearman18,teranishi2022evaluation,toda2022evaluation,umemoto2022location,ide2024interpretable,umemoto2023counterfactual,yeung2024strategic}. The stretch index, by contrast, reflects compactness through average player-to-centroid distances, thereby measuring the dispersion of defensive units \citep{clemente2015developing,bojinov2016collective,forcher2022defensive}. However, each metric has limitations when applied in isolation: pressure measures often overlook attacker interactions, while space control and stretch indices are too static to capture the fluid dynamics of transitions \citep{ogawa2025space}. This has motivated calls for context-sensitive, risk-weighted models that integrate spatial and temporal factors into defensive evaluation.

Parallel strands of research have also emphasized defensive positioning in specific game contexts, particularly crossing situations. The previous study \citep{Pafis2025Cross} demonstrated that the height of the defensive line and the origin of the cross significantly affect defensive success across elite teams in LaLiga, the Premier League, and the Bundesliga. Their findings highlight vulnerabilities when crosses originate from half-spaces, reinforcing the broader tactical emphasis on coordinated defensive positioning and subunit synchronization. These results align with those of \citet{Forcher2024_compactness}, who showed that local compactness near the ball rather than overall team compactness was the most robust predictor of defensive effectiveness. 

Despite recent advances in spatial metrics such as pitch control and expected possession value, these approaches primarily emphasize offensive progression. Existing defensive measures, by contrast, remain limited to isolated actions (e.g., tackles, interceptions) and fail to capture the coordinated role of the back four during transitions. There is still no framework that jointly evaluates the coordinated actions of the back-four line during negative transitions using synchronized tracking and event data. This gap limits both tactical understanding and the development of quantitative tools for coaches and analysts. Our study addresses this need by introducing interpretable spatio-temporal indicators of defensive line behavior, testing their predictive power, and demonstrating their tactical relevance through comparative analysis. In doing so, we provide a framework that connects defensive coordination to measurable outcomes, a necessary step for advancing both research and applied performance analysis.

Building on the tactical importance of defensive line coordination, this study aims to quantitatively evaluate how interpretable indicators of back-four organization relate to defensive outcomes during negative transitions. Specifically, we compare two elite teams namely FC Barcelona and Real Madrid CF, representing contrasting defensive philosophies.
We hypothesize that (1) Defensive success can be explained by measurable differences in spatial compactness, pressure intensity, and line positioning; (2) These relationships differ between teams due to distinct tactical behaviors.

\section*{Materials and methods}
\subsection*{Dataset}
%%%

% what data: tracking and event
In this study, we used two synchronized datasets: ($1$) tracking data from SkillCorner at $25$ Hz, providing positional and velocity information for all $22$ players and the ball, and ($2$) event data from StatsBomb, which includes detailed annotations of passes, duels, tackles, and pressures etc. \citep{hughes2019performance} for Laliga 2023/24 season.
% validity?
Despite being vision-based, SkillCorner has been validated in prior elite-level research for modeling defensive organization \citep{andrienko2022extracting}.

% almost all matches considered (one missing + why)
A total of 73 matches involving FC Barcelona and Real Madrid CF were analyzed, comprising 38 matches for Barcelona and 37 for Real Madrid (a Real madrid game data cannot be used). These matches were selected due to the teams' elite status and tactical diversity, offering a comprehensive dataset for examining collective defensive behavior. All of Barcelona's matches were included; however, one match for Real Madrid was excluded due to the absence of tracking data: specifically, the match between Valencia (Home) and Real Madrid (Away).

% synchronization
To achieve precise temporal alignment, we utilized ETSY (Event and Tracking data SYnchronization) \citep{van2023etsy}, a rule-based synchronization algorithm designed to correct time biases (e.g., kickoff shifts) and associate each event with the most plausible frame based on player and ball proximity, movement consistency, and physical plausibility \citep{van2023etsy}. For each successfully synchronized event, we extracted a $2.4$-second window centered on the matched frame ($30$ frames before and after) to examine defensive shape and player behavior during transitions. 

%%%

\subsection*{Preprocessing}

%%%

% sequence extraction + definition
After synchronization, tracking and event data were fully aligned, and a sequential pipeline extracted defensive sequences during negative transitions. 

% filtering out non relevant events
First, we systematically filtered non-possession events (substitutions, fouls, restarts) to isolate active open-play phases, subsequently identifying possession turnovers through changes in ball control between teams while retaining only those resulting from direct opponent interaction such as tackles, interceptions etc. rather than unforced errors. 
We excluded unforced errors (such as misplaced passes without defensive pressure, technical mistakes in open space, and goalkeeper distribution errors) to focus specifically on defensive effectiveness during opponent-induced turnovers rather than self-inflicted possession losses.

% normalization
Spatial coordinates were normalized to ensure consistent left-to-right attacking orientation across both match halves, with all transitions restricted to the defensive third (final $35$ meters).

% identification of 4 backs, gk, defensive third, possession player
For each transition, we algorithmically identified the defensive back-four by selecting four outfield players closest to their own goal line. This approach excluded goalkeepers and provided consistent defensive structure detection across varying team formations. 

% labeling of sequence
Defensive sequences were binary-labeled as successful ($1$) or failed ($0$) based on subsequent outcomes. Failures were defined as sequences resulting in opponent entry into the penalty area, shot attempts, or goals. All other sequences were labelled as success.

% temporal sequence
Finally, for each valid transition, we extracted a temporal sequence consisting of the $10$ subsequent events following the moment of possession loss to capture immediate defensive responses during critical transition moments. This event window length was chosen as it typically encompasses the key defensive behaviors and responses immediately following a turnover, such as pressures, tackles, and interceptions. Extending beyond $10$ events risks including offensive possessions or events not directly related to the defensive transition phase. While goals or critical incidents may occur outside this window (e.g., at the $12$th event), these are considered part of subsequent phases and do not impact the immediate defensive performance captured by the defined sequence. 

% Sequence number
This yielded $2,413$ high-quality transition sequences, comprising $1,434$ failed and $979$ successful defenses. Sequences were considered valid based on several criteria: ($1$) occurrence within the defensive third, defined as the ball being positioned at least $70$m from the team's own goal line, ($2$) the presence of a minimum of $4$ defenders within the frame, ($3$) the availability of complete tracking data for all relevant players, and ($4$) the absence of immediate restart events, such as throw-ins, offsides, or fouls, which could artificially truncate the defensive sequence. Our final dataset included $639$ defensive transition sequences for Barcelona with a success rate of $40.4\%$ and $624$ for Real Madrid with a success rate of $35.9\%$. 

%%%

\subsection*{Feature Engineering}
% Based on the motivation mentioned in the Introduction, here we describe the details of our proposed method.
% Feature 
To quantify defensive line behavior during negative transitions, we developed five interpretable rule-based metrics: \textit{Stretch Index}, \textit{Pressure Index}, \textit{Space Score}, \textit{Defense line height absolute}, and \textit{Defense line height relative to ball}. 

The relationship with the previous work is discussed in the discussion section.

% stretch index with fig reference
\subsubsection*{Stretch Index}

The Stretch Index measures the compactness and threat exposure of the last line of defense. It combines:
\begin{enumerate}
    \item the area of the convex hull formed by the four deepest defenders, and
    \item the mean distance from the three most advanced attackers to their nearest defender.
\end{enumerate}
This composite metric is calculated as:

\begin{equation}
\mathrm{StretchIndex}_{t} 
= \lambda \cdot \operatorname{ConvexHullArea}(\mathrm{Defenders}_{t}) 
+ (1-\lambda) \cdot \mathrm{PDA}_{t}
\label{eq:stretch_index}
\end{equation}
where $\lambda = 0.5$, and $\mathrm{PDA}_{t}$ denotes the Perceived Defensive Affordance, defined as the average shortest distance between each top attacker and the closest back-four defender. Here, \textit{t} denotes each frame index in the defensive sequence. The weighting parameter $\lambda = 0.5$ was chosen to give equal importance to spatial compactness (convex hull area) and attacker proximity (PDA).

% value interpretation
Lower values indicate tighter and better-coordinated lines whereas higher values indicate better defensive control of high-risk regions.

% pressure index
\subsubsection*{Pressure Index}

The Pressure Index reflects close marking intensity by counting attackers within a fixed radius of any defender:

\begin{equation}
\mathrm{PressureIndex}_{t} = 
\sum_{a \in \mathrm{Attackers}_{t}} 
\mathbb{I}\!\left(\min_{d \in \mathrm{Defenders}_{t}} \|a - d\| < r\right)
\label{eq:pressure_index}
\end{equation}
where $\mathbb{I}$ is the indicator function and radius $r = 3\,\mathrm{m}$. The value ranges from 0 to 3 where 0 means no attackers were marked and 3 means all attackers were marked within the range of interest. Higher values imply stronger pressure applied by defenders through tight marking.

% space score
\subsubsection*{Space Score}

\begin{figure}[ht]
\centering
\includegraphics[width=0.8\textwidth]{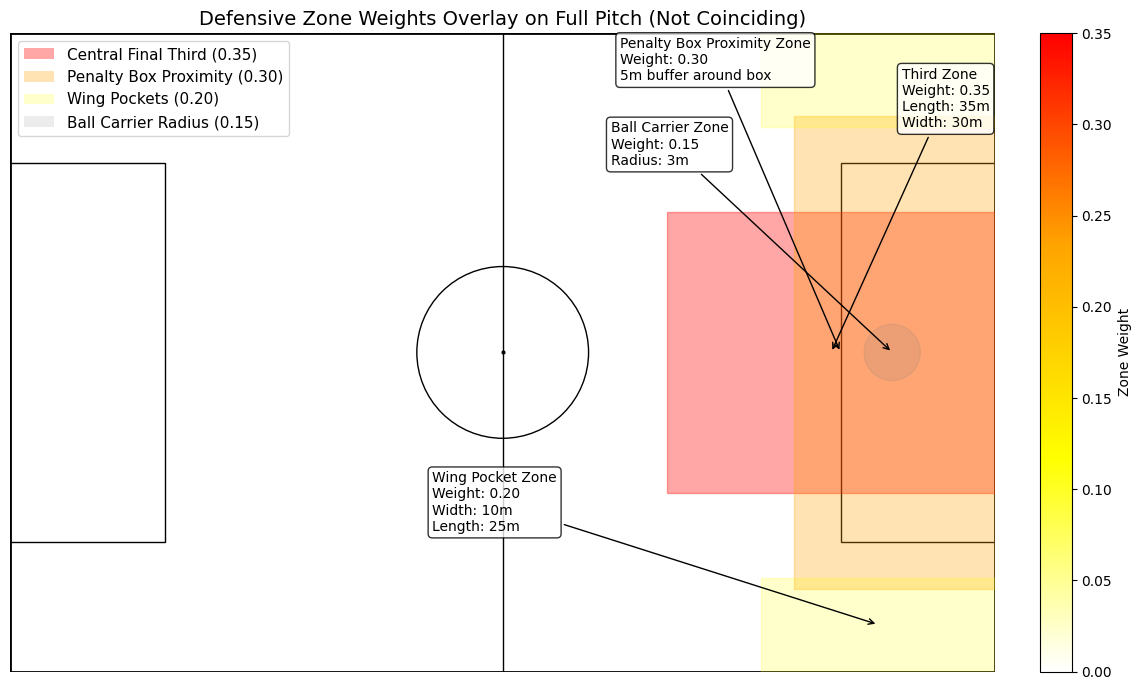}
\caption{Tactical zones used in the computation of the space score. The field is divided into four analytically defined zones prioritized by tactical importance: (1) Central Final Third (red), (2) Penalty Box Proximity (orange), (3) Wing Pockets (green), and (4) Ball-Carrier Radius (blue). Each zone is assigned a weight and dynamically assessed based on the number of defenders and attackers present. Zone overlaps are resolved by prioritizing higher-weighted zones. This framework enables frame-by-frame quantification of spatial control during defensive transitions. }
\label{fig:zonal distribution}
\end{figure}

The Space Score quantifies spatial dominance over tactically important zones. At each frame, we define four priority zones (see Table \ref{tab:zone_weights} and Figure \ref{fig:zonal distribution}): Central Final Third, Penalty Box Proximity, Wing Pockets, and Ball-Carrier Radius. Each zone $z$ has a tactical weight $w_z$, and the frame-level score is computed as:

\begin{equation}
\mathrm{SpaceScore}_{t} = \sum_{z \in Z} 
w_{z} \cdot \frac{D_{z}(t) - A_{z}(t)}{D_{z}(t) + A_{z}(t) + \epsilon}
\label{eq:space_score}
\end{equation}
where $D_{z}(t)$ and $A_{z}(t)$ are the numbers of defenders and attackers in zone $z$ at time $t$, and $\epsilon$ is a small constant ($\epsilon=1$) to avoid division by zero.

\begin{table}[ht]
\centering
\caption{Tactical Weights Assigned to Defensive Zones}
\label{tab:zone_weights}
\begin{tabular}{llc}
\toprule
\textbf{Zone} & \textbf{Definition} & \textbf{Weight ($w_z$)} \\
\midrule
Central Final Third & Area between box and center circle & $0.35$\\
Penalty Box Proximity & Buffer region outside penalty box & $0.30$\\
Wing Pockets & Lateral corridor inside final third & $0.20$\\
Ball-Carrier Radius & 5m radius around ball carrier & $0.15$\\
\bottomrule
\end{tabular}
\end{table}

Zone overlaps are resolved by assigning players to the highest-priority overlapping zone. A higher score indicates better defensive control of high-risk regions.

% Defense line height absolute
\subsubsection*{Absolute defensive line height}

Absolute defensive line height is defined as the positioning of the defensive line from the goal line. It is calculated as follows:

\begin{equation}
L_{t} = \frac{1}{4} \sum_{i \in D_{t}} x_{i,t}
\label{eq:absolute_height}
\end{equation}
where $x_{i,t}$ denotes the $x$-coordinate (field length direction) of defender $i$ at frame $t$. Note that $i \in D_{t}$ where $D_{t}$ is the set of the four deepest outfield defenders at time $t$.

% Defense line height relative to ball
\subsubsection*{Defensive line height relative to ball}

Relative line height with respect to ball quantifies the vertical coordination between the back line and the ball position. It is calculated as follows:

\begin{equation}
R_{t} = x^{\mathrm{ball}}_{t} - L_{t}
\label{eq:relative_height}
\end{equation}
where $x^{\mathrm{ball}}_{t}$ is the $x$-coordinate of the ball at frame t and $L_t$ is the absolute line height as defined in Eq.~\ref{eq:absolute_height}.

% one line comparison between absolute and relative height (why we created the feature)
The absolute line height $\overline{L}$ reflects the team's overall field coverage and risk appetite, with higher values indicating a more advanced defensive line whereas the relative measure $\overline{R}$ provides insight into the compactness and vertical coordination of the defensive unit with respect to the ball. Values of $\overline{R}$ near zero indicate that the defensive line is closely tracking the ball, while positive values indicate a deeper line relative to the ball's position.

\begin{figure}[htb]
  \centering
  \includegraphics[width=1\textwidth]{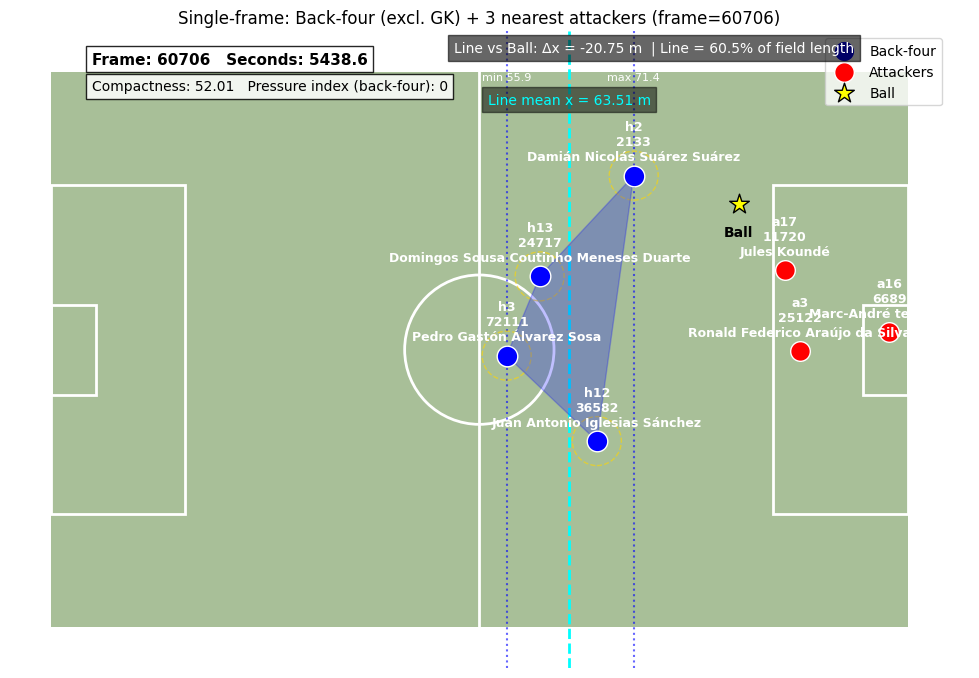}
  \caption{Single-frame illustration of back-four extraction and derived geometric features. The four deepest outfield defenders (blue markers) form the back-four polygon (shaded). Three nearest attackers are shown in red and the ball as a yellow star. Annotated values include frame ID and timestamp (top left), compactness (convex-hull area) and pressure index (top left box), the line mean (cyan dashed line) and the positions bounding the hull (blue dotted lines). This frame (example ID: 60706) is presented to clarify how per-frame measures (convex hull area, line mean, defender–attacker distances and ball-relative line height) are computed prior to sequence aggregation.}
  \label{fig:feature_visualization}
\end{figure}

To illustrate the feature computation process, Figure~\ref{fig:feature_visualization} presents a representative frame showing the back-four (blue markers), three nearest attackers (red markers), and the ball (yellow star). The defensive line hull, compactness, and line height relative to the ball were extracted from such configurations across all sequences.

% feature visualization (if not required, we can push to supplementary section)

% sanity check of features
\subsection*{Feature Aggregation and Selection}

To ensure interpretability and minimize redundancy, only mean values of each handcrafted feature were aggregated per defensive transition, rather than full statistical summaries (e.g., min, max, std), which tend to introduce multicollinearity in spatio-temporal data. Based on tactical relevance and prior multicollinearity analysis (see Supplementary Figure~\ref{S1_Fig}), the following five mean-based features were selected: $\mathrm{stretch\_index}$, $\mathrm{Pressure\_index}$, $\mathrm{Space\_score}$, $\mathrm{Defensive\_line\_height\_absolute}$, $\mathrm{Defensive\_line\_height\_relative}$.  

All features were standardized before statistical testing, and their raw ranges were examined to verify interpretability and absence of extreme outliers (Supplementary Table \ref{S1_Table}). 

\subsection*{Statistical Analysis}

To examine whether the engineered defensive indicators differed between successful and unsuccessful defensive outcomes across teams, a two-way ANOVA (Team × Outcome) was conducted for each of the five handcrafted features. The dependent variable was the mean feature value per defensive sequence, and the two factors were \textit{Team} (Barcelona, Real Madrid) and \textit{Defensive Outcome} (Success, Failure). 

ANOVA is a parametric statistical test used to assess whether the means of a quantitative dependent variable differ significantly across levels of one or more categorical independent variables. 

This analysis was performed using the Python package \texttt{pingouin} (v0.5.4).

For features showing significant Team × Outcome interactions ($p < 0.05$), post-hoc pairwise comparisons were performed using independent t-tests within each team to identify team-specific differences between successful and failed defensive transitions. To control for Type I error inflation within these post-hoc comparisons, Bonferroni correction was applied with a divisor of 4, yielding an adjusted significance threshold of $\alpha^* = 0.0125$ per comparison. 

% Post-hoc analysis strategy
For features demonstrating significant Team × Outcome interactions, within-team comparisons were conducted to examine whether defensive success versus failure differed significantly for each team independently. This approach allows identification of team-specific patterns in defensive behavior while maintaining statistical validity through appropriate alpha adjustment.

\subsection*{Machine Learning Prediction Models}

% \subsubsection{Objective}

% Purpose: Evaluate whether handcrafted defensive indicators can predict defensive success or failure.
Through predictive modeling we aim to validate whether handcrafted defensive indicators possess discriminative power for predicting defensive outcomes beyond their statistical significance demonstrated in the inferential analysis.
% why + bridge from ANOVA
While Two-way ANOVA established significant differences between successful and failed defensive sequences between teams, predictive modeling assessed the practical utility of these features for out-of-sample classification tasks.
This approach provides complementary validation: statistical significance indicates systematic differences, while predictive performance demonstrates practical applicability. The combination of interpretable handcrafted features with robust machine learning algorithms offers both tactical insights and empirical validation of defensive effectiveness metrics.

\subsubsection*{Data Preparation and team specific modeling}

We utilized the same five standardized mean defensive features employed in the statistical analysis:  Stretch\_index, Pressure\_index, Space\_score, Defensive\_line\_height\_absolute, Defensive\_line\_height\_relative.  

% team specific
Following the team-specific approach established in the ANOVA analysis, separate predictive models were developed for FC Barcelona and Real Madrid CF. This methodology recognizes that tactical systems and defensive philosophies differ substantially between teams, making team-specific models more appropriate than a unified approach. The dataset was filtered to include only sequences where possession was lost by Barcelona or Real Madrid, ensuring model focus on the defensive transitions of interest.

For each team, the dataset was split at the sample level into training (80\%) and test (20\%) subsets using stratified sampling to preserve the original class proportions. This ensured that both training and test data contained representative distributions of successful and failed defensive sequences.

\subsubsection*{Algorithm selection and model configuration}

% algorithms used + random seed 42
Based on the need for interpretable models capable of handling moderate-dimensional feature spaces, three primary algorithms were selected and compared. 

\textbf{Random Forest (RF)} \citep{ho1998random} is an ensemble method that provides feature importance scores and naturally handles non-linear relationships. The configuration included $300$ estimators, balanced class weighting, and a minimum of five samples per split. 

\textbf{XGBoost}\citep{Chen16} is a gradient boosting algorithm that offers superior performance on structured data with built-in mechanisms for class imbalance handling. The configuration employed $300$ estimators, a learning rate of $0.05$, maximum depth of three, and an optimized \texttt{scale\_pos\_weight} parameter. 

\textbf{Support Vector Classifier (SVC)}\citep{cortes1995support} is a non-linear classifier using the RBF kernel to capture complex decision boundaries. The configuration used balanced class weighting and enabled probability estimation for AUC calculation. 

All models used fixed random seeds (\texttt{seed=42}) to ensure reproducible results.

The primary approach employed built-in class weighting mechanisms: \texttt{class\_weight='balanced'} for Random Forest and Support Vector Classifiers, and \texttt{scale\_pos\_weight} parameter optimization for XGBoost. Additionally, an alternative upsampling strategy was tested using \texttt{sklearn.utils.resample} to balance the minority class during training while preserving the original test set distribution.

% Model Evaluation and Cross-validation Strategy
\subsubsection*{Model Validation}

Model performance was assessed using a rigorous evaluation framework combining train--test validation with cross-validation. Again, each team's dataset was split into $80$\% training and $20$\% testing sets for each team to preserve class distributions.

\textbf{Performance Metrics:} Model performance was evaluated using multiple complementary metrics: ROC-AUC, accuracy, precision, recall, and F1-score. ROC-AUC was chosen as the primary metric because it
%it is robust to moderate class imbalance and
measures the model’s ability to discriminate between successful and failed defensive transitions across all possible classification thresholds. Accuracy, precision, recall, and F1-score provide additional insights into practical classification performance. 

\textbf{Cross-Validation:} Five-fold stratified cross-validation was applied to the training set for both model selection and performance estimation, ensuring each fold maintained the original class distribution. 

\subsubsection*{Feature Importance}

Feature importance was evaluated to interpret model behavior and identify the most influential defensive indicators contributing to predictive performance. For tree-based models (Random Forest and XGBoost), built-in importance scores were obtained based on the mean decrease in impurity. In addition, SHAP (SHapley Additive Explanations) values \citep{Lundberg2017SHAP} were computed for the best-performing classifier to provide a model-agnostic and locally interpretable assessment of feature contributions.

This combined approach ensures interpretability across different model types (tree-based and kernel-based) and allows comparison between global model-derived importances and local SHAP explanations. Specific feature rankings and their tactical implications are reported in the Results section.
In addition to SHAP, the built-in feature importance scores of the corresponding tree-based model were compared to the SHAP-derived rankings to confirm stability of the importance ordering across interpretability methods. Both approaches consistently identified \textit{space score}, \textit{relative line height}, and \textit{stretch index} as the most influential features in predicting defensive success.

All analyses were implemented in Python 3.x using \texttt{scikit-learn} (v1.9.3) for machine learning algorithms, \texttt{XGBoost} (v2.0) for gradient boosting, and \texttt{SHAP} (v0.45) for model interpretability. Statistical preprocessing employed \texttt{pandas} (v1.5) and \texttt{NumPy} (v1.24) for data manipulation, with \texttt{matplotlib} and \texttt{seaborn} for visualization.

\section*{Results}

\subsection*{Statistical Analysis for each feature}

Two-way ANOVA revealed significant main effects and interactions across multiple defensive features (Table~\ref{tab:anova_results}). 

Among these, \textbf{line height (relative)} exhibited the highest F-value ($F=430.06$, $p<0.001$, $\eta^2_p=0.153$), indicating that the defensive line’s vertical position relative to the ball was the most discriminative factor between successful and unsuccessful defensive outcomes.   

% team effect first (hypothesis 1)
A significant effect of \textit{Team} was observed for all five indicators  ($p<0.01$), indicating systematic differences in defensive configurations between FC Barcelona and Real Madrid CF.

% outcome effect (hypothesis 2)
The main effect of \textit{Defensive Outcome} (Success vs Failure)  was also significant for all features ($p<0.05$), suggesting that successful defensive transitions were characterized by distinct spatial and pressure dynamics. 

% Interaction effect ((hypothesis 3)
Significant \textit{interaction effects} between \textit{team lost possession} and \textit{label} were observed for \textbf{pressure} ($F=1.997$, $p=0.006$, $\eta^2_p=0.016$), \textbf{stretch} ($F=2.638$, $p<0.001$, $\eta^2_p=0.021$), and \textbf{line height (relative)} ($F=2.200$, $p=0.002$, $\eta^2_p=0.017$). These interactions suggest that the relationship between defensive configuration and success varied between Barcelona and Real Madrid.
\begin{table}[htbp]
\centering
\caption{Two-way ANOVA Results for Defensive Features (Team $\times$ Outcome)}
\label{tab:anova_results}
\small
\begin{tabular}{llrrr}
\hline
\textbf{Feature} & \textbf{Source} & \textbf{F} & \textbf{p (unc.)} & $\eta^2_p$ \\
\hline
Space Score & Team & 3.031 & $<.001$ & .024 \\
Space Score & Outcome & 23.882 & $<.001^{***}$ & .010 \\
Space Score & Team $\times$ Outcome & 0.648 & .871 & .005 \\
Pressure Index & Team & 4.430 & $<.001^{***}$ & .034 \\
Pressure Index & Outcome & 26.907 & $<.001^{***}$ & .011 \\
Pressure Index & Team $\times$ Outcome & 1.997 & .006$^{**}$ & .016 \\
Stretch Index & Team & 3.707 & $<.001^{***}$ & .029 \\
Stretch Index & Outcome & 47.864 & $<.001^{***}$ & .020 \\
Stretch Index & Team $\times$ Outcome & 2.638 & $<.001^{***}$ & .021 \\
Line Height (Relative) & Team & 4.774 & $<.001^{***}$ & .037 \\
Line Height (Relative) & Outcome & 430.063 & $<.001^{***}$ & .153 \\
Line Height (Relative) & Team $\times$ Outcome & 2.200 & .002$^{**}$ & .017 \\
Line Height (Absolute) & Team & 2.215 & .002$^{**}$ & .017 \\
Line Height (Absolute) & Outcome & 5.110 & .024$^{*}$ & .002 \\
Line Height (Absolute) & Team $\times$ Outcome & 0.527 & .952 & .004 \\
\hline
\end{tabular}

\vspace{4pt}
\par\noindent\footnotesize
Significance levels: * ($p<0.05$), ** ($p<0.01$), *** ($p<0.001$). \\
$F$ = Fisher's F-statistic; $\eta^2_p$ = partial eta-squared (effect size).
\end{table}

\begin{figure}[ht]
\centering
\includegraphics[width=0.8\linewidth]{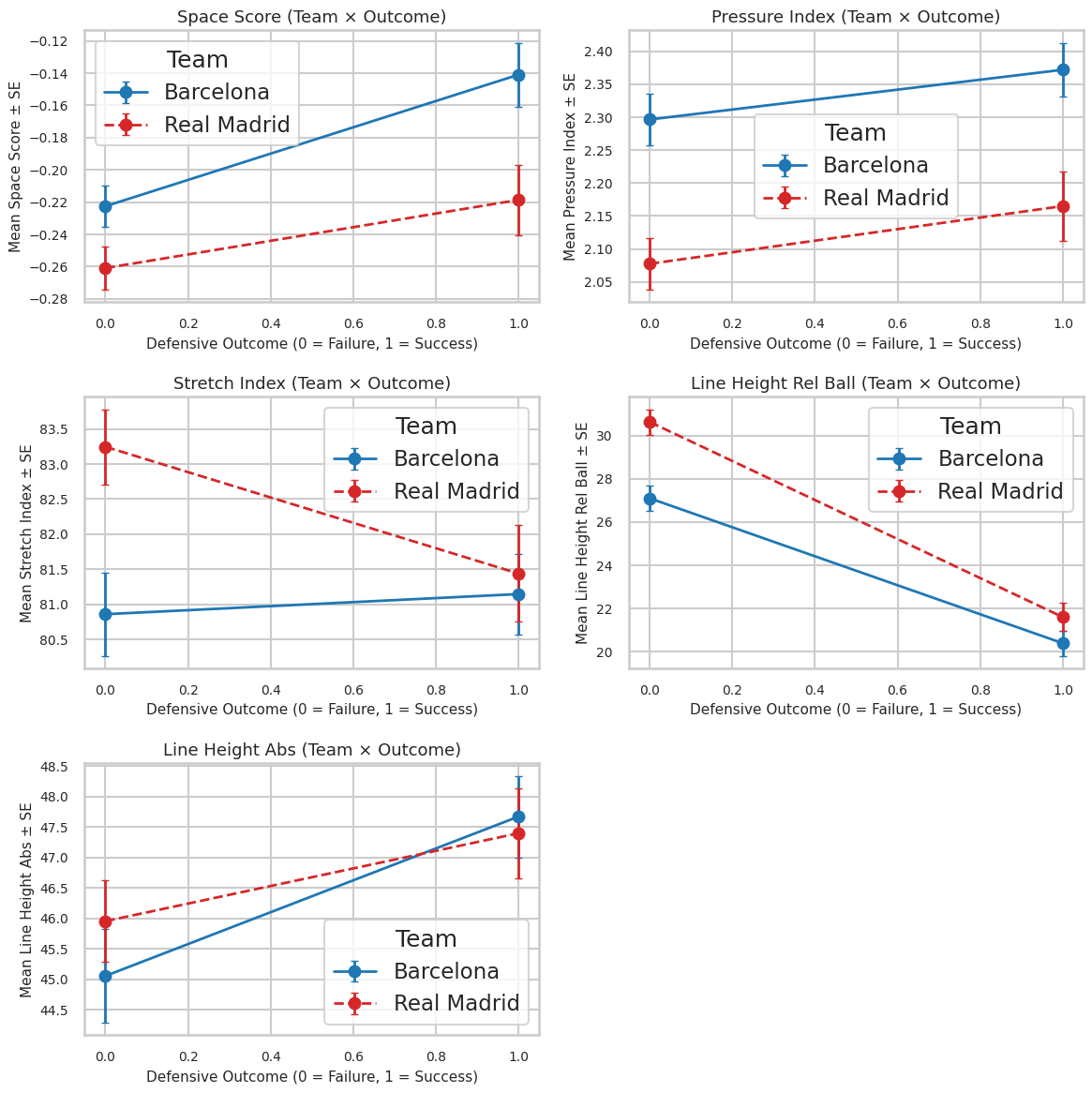}
\caption{Interaction effects between team and defensive success across features. Line height (relative) exhibits a clear crossover pattern, indicating differential team behavior during transitions.}
\label{fig:interaction}
\end{figure}
As illustrated in Figure~\ref{fig:interaction}, \textbf{Barcelona} maintained a consistently higher defensive line during successful defensive phases, while \textbf{Real Madrid} showed a steeper decline in line height during unsuccessful outcomes—indicating greater sensitivity to transition dynamics. Differences in \textbf{pressure} and \textbf{stretch} were more subtle, although Real Madrid displayed slightly higher variability in pressure index.

To account for post-hoc comparisons, post-hoc corrections were applied using Bonferroni (Table~\ref{tab:alpha_correction}). After correction, only \textbf{line height (relative)} remained significant for both teams (Bonferroni $p<.001$), confirming its strong association with defensive success. For Real Madrid, \textbf{stretch index} reached marginal significance under FDR correction ($p=.081$), while other features were not significant under any correction method.

\begin{table}[htbp]
\centering
\caption{Post-hoc multiple testing correction for selected features.}
\label{tab:alpha_correction}
\small
\begin{tabular}{llrr}
\hline
\textbf{Team} & \textbf{Feature} & \textbf{p (orig.)} & \textbf{p (Bonf.)} \\
\hline
Barcelona & Pressure Index & .196 & 1.000 \\
Barcelona & Stretch Index & .741 & 1.000 \\
Barcelona & Line Height (Rel.) & $<.001$ & $<.001^{***}$ \\
\hline
Real Madrid & Pressure Index & .180 & 1.000 \\
Real Madrid & Stretch Index & .041 & .244 \\
Real Madrid & Line Height (Rel.) & $<.001$ & $<.001^{***}$ \\
\hline
\end{tabular}

\vspace{4pt}
\par\noindent\footnotesize
Significance levels: * ($p<0.05$), *** ($p<0.001$). \\
ns = not significant after correction.
\end{table}

% \subsubsection*{Feature selection summary}
In summary, for two-way ANOVA and post-hoc comparison, \textbf{line height (relative to the ball)} was retained as the most robust predictor of defensive success across both teams. For Real Madrid, \textbf{stretch index} showed a weak but noteworthy trend (FDR $p=.081$), suggesting potential tactical variability under transition conditions. Other features such as \textbf{pressure index} and \textbf{absolute line height} were excluded from subsequent modeling due to non-significant corrected effects.

\subsection*{Predictive Modeling Results}

%%%
To assess whether handcrafted defensive indicators can predict defensive success, we trained and validated multiple supervised learning models for each team. 

\subsubsection*{Model Performance and Comparison}
%%%
Table~\ref{tab:model_performance} summarizes the predictive performance of three algorithms across both teams.

\begin{table}[ht]
\centering
% \begin{threeparttable}
\caption{Predictive model performance by team. Ensemble models outperform linear and distance-based baselines, suggesting non-linear dependencies among handcrafted defensive indicators.}
\label{tab:model_performance}
\begin{tabular}{llrrrrr}
\toprule
\textbf{Team} & \textbf{Algorithm} & \textbf{ROC AUC} & \textbf{Accuracy} & \textbf{Precision} & \textbf{Recall} & \textbf{F1-Score} \\
\midrule
\multirow{3}{*}{Barcelona} 
 & \textbf{XGBoost} & \textbf{0.724} & 0.672 & \textbf{0.683} & 0.645 & 0.663 \\
 & Random Forest & 0.718 & 0.664 & 0.671 & 0.652 & 0.661 \\
 & SVM (RBF) & 0.689 & 0.634 & 0.645 & 0.618 & 0.631 \\
\midrule
\multirow{3}{*}{Real Madrid} 
 & \textbf{XGBoost} & \textbf{0.698} & 0.648 & \textbf{0.659} & 0.634 & 0.646 \\
 & Random Forest & 0.693 & 0.642 & 0.651 & 0.628 & 0.639 \\
 & SVM (RBF) & 0.664 & 0.618 & 0.627 & 0.605 & 0.616 \\
\bottomrule
\end{tabular}%
\begin{tablenotes}
\small
\item Best-performing models per team are shown in \textbf{bold}.
\end{tablenotes}
%\end{threeparttable}
\end{table}
Three supervised learning models XGBoost, Random Forest, and Support Vector Machine (SVM) were trained separately for each team to predict defensive success from handcrafted indicators. All models used standardized input features, an 80–20 stratified train–test split, and identical cross-validation procedures to ensure fair comparison.

As shown in Table~\ref{tab:model_performance}, XGBoost achieved the highest discriminative performance for both teams (Barcelona: ROC AUC = 0.724; Real Madrid: ROC AUC = 0.698), followed closely by Random Forest. The SVM with an RBF kernel produced lower accuracy and precision, indicating that the handcrafted indicators exhibit non-linear interactions better captured by ensemble tree-based models.

Across all models, ROC-AUC values ranged between $0.60$ and $0.72$, confirming that handcrafted features capture meaningful variance related to defensive success. 

Team-level performance differences indicate that the handcrafted indicators are slightly more predictive for Barcelona than for Real Madrid. This finding aligns with the inferential results, where Barcelona’s defensive outcomes showed stronger feature–outcome relationships (e.g., Space Score, Relative Line Height). In contrast, Real Madrid’s lower classification performance suggests more variable or less structured defensive responses during transitions.

Although all three classifiers showed reasonable performance, XGBoost provided the most consistent balance between sensitivity and specificity. Consequently, it was selected for subsequent model interpretability using SHAP values. This choice ensures that feature attribution reflects the best-performing predictive framework while maintaining transparency in model behavior.

%%%

%%%

\textbf{Reducing false positives: a precision-focused evaluation.}  
In defensive prediction, a false positive corresponds to incorrectly classifying a failed defense as successful—an error that may yield misleading tactical conclusions. 
Hence, \textit{precision} (the proportion of correctly predicted successes among all predicted successes) was prioritized, as it directly penalizes false positives.
XGBoost achieved the highest precision for both teams (0.683 for Barcelona, 0.659 for Real Madrid), confirming its ability to minimize false alarms while maintaining good recall.

\textbf{Rationale for XGBoost as the main model.}  
XGBoost was chosen for further interpretation due to its:
\begin{enumerate}
    \item superior balance between ROC-AUC and precision across both teams,
    % \item built-in handling of class imbalance via \texttt{scale\_pos\_weight},
    \item robustness to correlated spatial indicators, and
    \item interpretability through SHAP analysis.
\end{enumerate}

\subsubsection*{Feature Importance and SHAP Interpretation}

Feature importance analysis and SHAP attribution consistently identified \textbf{Space Score} and \textbf{Relative Line Height} as the strongest predictors of defensive success (Table~\ref{tab:feature_importance}). These features jointly capture the ability to control high-risk zones and maintain vertical synchronization with the ball—both critical to successful transitions.

\begin{table}[ht]
\centering
\caption{Feature importance rankings by team and method. Higher scores indicate greater contribution to defensive success prediction.}
\label{tab:feature_importance}
\begin{tabular}{l l r r r r}
\toprule
\textbf{Rank} & \textbf{Feature} & 
\multicolumn{2}{c}{\textbf{Barcelona}} & 
\multicolumn{2}{c}{\textbf{Real Madrid}} \\
\cmidrule(lr){3-4} \cmidrule(lr){5-6}
 &  & \textbf{Feature Importance} & \textbf{SHAP} & \textbf{Feature Importance} & \textbf{SHAP} \\
\midrule
1 & Space Score & 0.347 & 0.312 & 0.329 & 0.298 \\
2 & Line Height (Relative) & 0.251 & 0.267 & 0.243 & 0.259 \\
3 & Stretch Index & 0.198 & 0.203 & 0.187 & 0.196 \\
4 & Pressure Index & 0.112 & 0.118 & 0.127 & 0.134 \\
5 & Line Height (Absolute) & 0.092 & 0.100 & 0.114 & 0.113 \\
\bottomrule
\end{tabular}
\end{table}
The consistency between SHAP-derived feature importance (Table \ref{tab:feature_importance}) and effect sizes ($\eta^2$) from the ANOVA (Table \ref{tab:anova_results}) reinforces the interpretability of the results. Specifically, features showing significant Team × Outcome interactions in Table \ref{tab:anova_results}, such as \textit{Space Score} and \textit{Relative Line Height} also ranked highest in SHAP importance (Table \ref{tab:feature_importance}), validating their explanatory and predictive utility. 

\subsubsection*{Integration of Statistical and Predictive Model Findings}

The convergence between inferential and predictive analyses strengthens the interpretability and reliability of the handcrafted indicators. 
While ANOVA identified \textit{relative line height} and \textit{space control} as statistically discriminative, predictive modeling confirmed their practical predictive power.
Barcelona’s higher model precision and consistency reflect a more structured and predictable defensive system, whereas Real Madrid’s slightly lower and more variable scores suggest adaptive but less stable transition behaviors.

Overall, these results demonstrate that interpretable spatio-temporal indicators can both explain and predict defensive outcomes thus, bridging the gap between descriptive analytics and practical decision-support in elite soccer.

\section*{Discussion}

% main findings
This study quantitatively examined the collective behavior of the back-four defensive line using interpretable spatio-temporal indicators derived from synchronized tracking and event data. By integrating inferential statistics with predictive modeling, we identified the key spatial mechanisms underlying defensive success during negative transitions.
In this section, we discuss the results, methodology, limitation, and future work. 

% main results first
Across analytic layers, \textbf{relative line height} emerged as the most robust indicator of defensive success. It produced the largest ANOVA effect (F = $430.06$, $p<0.001$, $\eta^2_p = 0.153$), survived Bonferroni-corrected post-hoc comparisons for both teams, and ranked among the top predictors in SHAP-based feature attribution. \textbf{Space score} also showed consistent importance: it exhibited significant effects in the ANOVA and was highly ranked in SHAP, indicating that defensive control over high-risk zones (central final third and penalty proximity) is both statistically and predictively meaningful.
Other features showed more nuanced patterns. 
\textbf{Pressure index} produced significant Team × Outcome effects in the ANOVA but did not survive Bonferroni-adjusted post-hoc testing, indicating a detectable but not robust within-team difference after conservative correction. 
Conversely, \textbf{stretch index} did not demonstrate Bonferroni-significant within-team differences, yet it contributed meaningfully to model predictions (moderate SHAP importance). 
The divergence of statistical significance in ANOVA (interaction-level signal) versus predictive importance suggests that stretch captures nonlinear or context-dependent compactness properties that are exploited by tree-based models but are not fully captured by marginal linear contrasts.
These results support a tactical interpretation in which vertical coordination relative to the ball (relative line height) and dominance of high-value zones (space score) are central to preventing penetration and dangerous entries during transitions. 
Higher, synchronized lines relative to the ball were associated with successful recoveries and reduced opponent access to penalty-area proximities, consistent with pressing and coordinated recovery principles. Pressure index and absolute line height appear complementary: they describe local marking intensity and overall depth, respectively, but are less reliable as standalone predictors than coordination-focused metrics.

Team-specific patterns are evident. 
% team difference
Barcelona showed stronger and more consistent feature–outcome relationships and higher model predictability, implying that their back-four behaviours are more systematic and thus more readily captured by global summary features. 
Real Madrid exhibited weaker and more variable relationships, suggesting a more adaptive or context-driven defensive approach where the same static summary may only partially capture successful behaviours.
These differences demonstrate how team philosophy manifests statistically and underscore the value of combining interpretable metrics with team-specific analysis.
Collectively, these results indicate that compactness, spatial control, and vertical coordination relative to the ball are key determinants of defensive success during negative transitions.

% tactical impact
As tactical impact, the proposed metrics provide a data-driven framework for evaluating and training defensive organization.  
For coaches, the \textit{Relative Line Height} and \textit{Space Score} can serve as interpretable indicators for monitoring compactness and spatial control in real time.  
By visualizing these indicators post-match or during training, staff can identify phases where the line becomes excessively deep or spatially stretched, enabling targeted feedback.  
Importantly, team-specific interpretations are essential—what constitutes optimal compactness for Barcelona may differ from Real Madrid’s flexible transition style.

%\subsection*{Methodological Considerations}
Methodologically, using a fixed 10-event sequence window allowed consistent evaluation of immediate post-turnover defensive responses but may overlook longer recovery sequences.  
Feature aggregation via mean values enhanced interpretability and reduced multicollinearity but sacrifices temporal granularity.  
Similarly, predictive modeling relied solely on handcrafted defensive indicators, excluding contextual information such as opponent actions, ball trajectory, or pitch zones outside the defensive third, which may further influence success probability. The exclusion of one Real Madrid match due to missing tracking data is unlikely to affect the overall findings, as the remaining sequences adequately represent the team's defensive behavior across the season.

% comparison with previous research

Our findings are broadly aligned with prior literature emphasizing compactness and spatial control in defensive performance \citep{forcher2022defensive,clemente2015developing} where earlier studies typically measured global compactness or used single-snapshot metrics, we demonstrate that line height relative to ball measures and zone-weighted space control provide superior discriminative and predictive power during transitions. This extends work on pitch control and zonal influence \citep{Spearman18,teranishi2022evaluation} into the defensive domain and corroborates recent evidence that line height and pressing timing meaningfully affect defensive success \citep{Pafis2025Cross,Forcher2024_compactness}.

Importantly, the joint ANOVA + SHAP framework highlights complementarities and limits of prior approaches. For example, stretch/compactness metrics historically reported as important can show only modest marginal effects but nevertheless contribute in nonlinear combinations—an effect that conventional linear inference may underestimate. Our approach therefore reconciles inferential and predictive perspectives: inference identifies robust, generalizable contrasts (e.g., relative line height), while predictive models reveal higher-order interactions (e.g., stretch interacting with attacker configuration) that can improve out-of-sample discrimination.

%%%
% For comparison with previous research, our findings extend prior research on defensive compactness \cite{forcher2022defensive,clemente2015developing} and pitch control \cite{Spearman18,teranishi2022evaluation} by demonstrating that interpretable handcrafted indicators can quantify coordination dynamics during transitions.  
% The significant role of relative line height aligns with studies linking advanced defensive positioning to effective pressing \cite{Pafis2025Cross,Forcher2024_compactness}.  
% However, while prior metrics emphasized global compactness, our results highlight that context-sensitive, ball-relative measures provide more predictive and interpretable insights into defensive effectiveness.

%limitation
For the limitation of this study, the analysis was limited to two teams from a single league (LaLiga 2023/24), restricting generalizability.  
Vision-based tracking may introduce minor spatial inaccuracies, and one Real Madrid match was excluded due to data loss.  
Furthermore, the study focuses exclusively on back-four systems; its applicability to back-three or hybrid formations remains to be verified.  
Finally, defensive behavior is inherently context-dependent, and factors such as opponent quality, fatigue, or tactical match state were not explicitly modeled.

% future direction
Future work should extend this framework across multiple leagues and tactical systems to assess generalizability.  
Incorporating dynamic attacker–defender interactions using temporal graph neural networks or probabilistic modeling could capture the evolving structure of defensive coordination.  
Additional contextual variables such as ball speed, field location, and opponent strength should be integrated to enhance predictive robustness.  
Lastly, embedding these interpretable indicators into real-time analytical systems could support in-game tactical decision-making and training feedback loops.

% conclusion
In conclusion, this study demonstrates that interpretable handcrafted spatial features can effectively characterize and predict defensive performance in elite soccer.  
By combining statistical inference and predictive modeling, we bridge the gap between tactical interpretation and empirical validation.  
The integration of \textit{relative line height}, \textit{space control}, and \textit{compactness metrics} provides a coherent framework for quantifying collective defensive behavior—advancing both analytical research and practical performance analysis in the modern game.

\backmatter

\ifarxiv
\bmhead{Acknowledgments}
This work was financially supported by JSPS KAKENHI Grant Number 23H03282. 
\fi

\section*{Declarations}
\begin{itemize}
%\item Funding: This work was financially supported by JST SPRING  Grant Number JPMJSP2125 and JSPS KAKENHI Grant Number 23H03282.
\item Competing interests: The authors have no competing interests to declare that are relevant to the content of this article.
%\item Ethics approval: Not applicable
%\item Consent to participate: Not applicable
%\item Consent for publication: Not applicable
% \item Availability of data and materials: The data of the research is publicly available with details in \url{https://github.com/statsbomb/statsbombpy}

% \item Code availability: The code for the model is available at
% % \ifarxiv
% % \url{https://github.com/calvinyeungck/Football-Match-Event-Forecast}.
% % \else
% \url{https://github.com/calvinyeungck/Analyzing-Two-Agents-Interaction-in-Football-Shot-Taking-Situations}.
% \fi
\end{itemize}

\newpage
\bibliography{sn-article}% common bib file
% \input output.bbl

\begin{appendices}
\section*{Suppelementary Materials}

\begin{figure}[ht]
\centering
\includegraphics[width=0.8\textwidth]{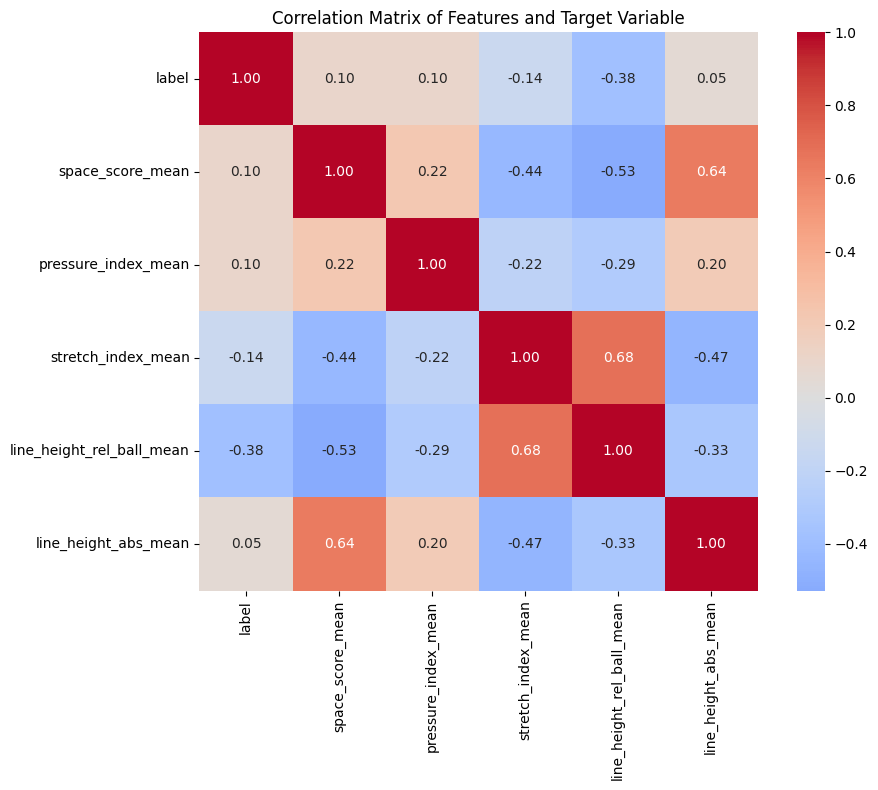}
\caption{Correlation heatmap of handcrafted defensive indicators.
While some moderate correlations exist (e.g., line height and space score),
multicollinearity remains acceptable across features.}
\label{S1_Fig}
\end{figure}

\begin{table}[ht]
\centering

\caption{\textbf{Outlier Assessment and Data Integrity.}
(A) Descriptive summary of raw defensive indicators and detected outliers before conducting inferential or predictive analyses.
(B) Outlier detection summary across 2,413 defensive sequences.}
\vspace{0.5em}

% ---------- PART (A) ----------
\textbf{(A) Descriptive summary of raw defensive indicators}\\[0.3em]
\begin{tabular}{lcccccccc}
\toprule
\textbf{Feature} & \textbf{Min} & \textbf{Q1} & \textbf{Median} & \textbf{Q3} & \textbf{Max} & \textbf{Mean} & \textbf{SD} & \textbf{Range} \\
\midrule
Space Score & -0.905 & -0.440 & -0.291 & -0.093 & 0.730 & -0.237 & 0.294 & 1.635 \\
Pressure Index & 0.000 & 2.000 & 2.000 & 3.000 & 3.000 & 2.135 & 0.785 & 3.000 \\
Stretch Index & 37.369 & 76.664 & 84.097 & 89.121 & 106.312 & 81.329 & 11.352 & 68.943 \\
Line Height (Relative) & -14.476 & 18.390 & 27.577 & 36.629 & 56.189 & 26.926 & 12.813 & 70.665 \\
Line Height (Absolute) & 11.640 & 39.528 & 44.998 & 52.970 & 88.486 & 46.947 & 12.710 & 76.846 \\
\bottomrule
\end{tabular}

\vspace{1em}

% ---------- PART (B) ----------
\textbf{(B) Outlier detection summary across 2,413 defensive sequences}\\[0.3em]
\begin{tabular}{lcccc}

\toprule
\textbf{Feature} & \textbf{z-score Outliers} & \textbf{IQR Outliers} & \textbf{Lower Bound} & \textbf{Upper Bound} \\
\midrule
Space Score  & 9 & 73 & -0.962 & 0.428 \\
Pressure Index  & 0 & 51 & 0.500 & 4.500 \\
Stretch Index  & 25 & 136 & 57.980 & 107.805 \\
Line Height (Relative) & 2 & 8 & -8.968 & 63.987 \\
Line Height (Absolute)  & 9 & 112 & 19.366 & 77.731 \\
\bottomrule
\end{tabular}

\vspace{0.75em}
\noindent\textit{Outlier rates remained below 6\% for all features under both detection methods. Visual inspection revealed no systematic skewness or clustering of extreme values, confirming that all data points were retained for subsequent analyses.}
\label{S1_Table}
\end{table}

\end{appendices}

\end{document}